\documentclass[%
 reprint,
showpacs,preprintnumbers,
 amsmath,amssymb,
 aps,
]{revtex4-1}

\usepackage{graphicx}
\usepackage{dcolumn}
\usepackage{bm}

\begin{document}


\title{Magnetoresistance of double layer hybrid system in tilted magnetic field}
\author{Fariborz Parhizgar}
\author{Reza Asgari}
 \email{asgari@ipm.ir}
\affiliation{School of Physics, Institute for Research in Fundamental Sciences (IPM), Tehran 19395-5531, Iran}

\date{\today}

\begin{abstract}

The magnetoresistance and Hall coefficient of a doped graphene layer are investigated in the presence of a tilted magnetic field. We consider a graphene layer assembled by either another graphene layer or by a two-dimensional electron gas (2DEG) layer and with the interlayer electron-electron interaction modeled within the random phase approximation. Our calculated magnetoresistances show different interlayer screening effects between decoupled graphene-graphene and graphene-2DEG systems. We also analyze the dependence of dielectric materials as well as the distance between the layers on magnetoresistances. The angle dependence of the Hall coefficient is studied and we show that a quite large Hall resistivity occurs in the graphene layer.
\end{abstract}

\maketitle

\section{Introduction}

Magnetoresistance~\cite{Nagaev}, the change of a material's resistivity in the presence of an external magnetic field, has been of interest both as a tool to probe the fundamental properties of an electronic material and for technological applications. Classically, the magnetoresistivity effect depends on both the
strength of the magnetic field and the relative direction of the magnetic field with respect to the current due to the Lorentz force. For nonmagnetic metals, magnetoresistivity effects at low magnetic fields are very small, although the effect can become quite large for high magnetic fields.

The magnetoresistance of graphene, a one atom thick layer of carbon atoms arranged in a honeycomb lattice~\cite{Novoselov,CastroNeto}, in the presence of the in-plane magnetic field has been studied by Hwang and Das Sarma by using the Boltzmann approach~\cite{ref:dassarma1}.
In the presence of an in-plane magnetic field, charge carriers of graphene are spin polarized and the effect of the magnetic vector potential is negligible owing to the one atom thickness of graphene. The authors showed that the applied magnetic field gives rise to the increase of the resistivity of graphene up to a saturation field where all electrons of the conduction band are spin polarized. The magnetic field beyond the saturation field excites electrons from the valence band to the conduction band and leads to a negative differential magnetoresistance.
In practice, it is difficult to reach the saturation magnetic field of a graphene layer, $B_{S}\sim 140\sqrt{\bar{n}}$ where $\bar{n}$ is in units of $10^{10}$cm$^{-2}$ and $B_S$ is scaled by units of Tesla. Although using the low electron density of graphene is feasible in experiments~\cite{ref:Geim}, one has to use an extremely clean and pristine sample to meet the necessary condition in which impurity density is much less than the charge carrier density to avoid any localization regime.

The magnetoresistance properties of a graphene layer are in contrast to those obtained in conventional two-dimensional electron gas (2DEG) systems in which the resistivity increases up to a certain magnetic field and then saturates~\cite{2d-exp, 2DEG-exp, 2DEG-theory,Herbut}. The magnetoresistance behavior of the 2DEG system can be understood by the Zeeman coupling and the reduction of screening of charge impurities in a polarized Fermi liquid system.

If the magnetic field has a small deviation from the sample plane, in addition to the spin polarization of conduction electrons, it gives rise to the Hall effect. Tilting the magnetic field has been shown to be a straightforward tool to disentangled spin and orbit effects in 2DEG systems~\cite{Kuntsevich}. The longitudinal resistivity and Hall coefficient of a conventional 2DEG have been investigated experimentally~\cite{Vitkalov} and theoretically~\cite{Herbut} in a slightly tilted magnetic field.

Assembling graphene, on the other hand, with various 2D layers into artificial heterostructures to explore novel or tailed properties has been proposed~\cite{novoselov-12} and realized in tunneling effect transistors~\cite{dean-12}. Hybridizing a gapless graphene layer with another gapless graphene layer makes the system a decoupled layer graphene (DLG)~\cite{Profumo DLG} where both layers are chiral and differs from a situation in which a gapless graphene is assembled by a 2DEG sample (G-2DEG)~\cite{Principi Hybrid} which is a chiral-nonchiral hybrid system. This structure makes it possible to study transport properties of the graphene layer according to an interlayer electron-electron interaction.

In this paper, we consider a hybrid structure in the presence of a tilted magnetic field to explore the effect of the interlayer interaction on the magnetoresistance properties of the studied layer. The longitudinal resistivity and Hall coefficient of the studied graphene layer are explored and the results are compared with those results obtained for only a single layer graphene in the presence of a magnetic field. We also analyze the dependence of the dielectric material which  fills the space between the two layers as well as the distance between layers on the magnetoresistance and Hall coefficient and show that the interlayer interaction plays a vital role even at longer distances by using a strong dielectric material between the two layers. The angle dependence of the Hall coefficient is obtained and we show that a quite large Hall resistivity occurs in graphene for certain values of the carrier density and screened interaction. This particular result is in contrast to that obtained in a 2DEG system where the Hall coefficient increases up to $30 \%$.

The paper is organized as follows. In Sec.~\ref{sec2}, we describe a model Hamiltonian of double layer systems and then derive the conductivity and Hall coefficient of the studied layer in a tilted magnetic field. Section~\ref{sec3} is devoted to our numerical results of the longitudinal conductivity and Hall coefficient of the studied graphene layer in a hybrid structure. Finally, a brief summary of
results is given in Sec.~\ref{sec4}.

\section{Model and Theory}\label{sec2}

\begin{figure}
\includegraphics[width=1.0\linewidth]{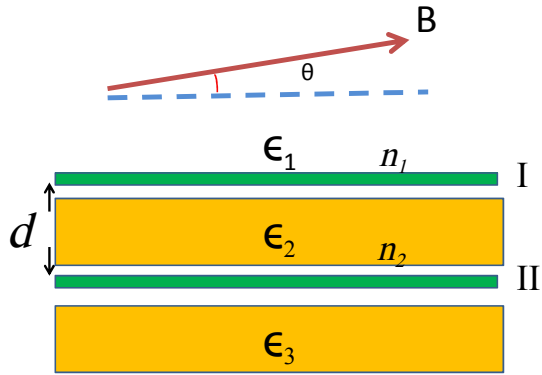}
\caption{(Color online) A schematic of a decoupled bilayer 2D system in the presence of a tilted magnetic field. In our study, the first layer is always graphene (studied graphene layer) with charge density $n_1$ and the second layer (layer II) is either a graphene layer or a 2DEG layer which is separated by a distance $d$ from the studied graphene layer. The materials in the hybrid structure are indicated by their dielectric constants.
\label{fig:schem}}
\end{figure}

We consider a double layer structure incorporating a dopped graphene layer (layer I) placed on another two-dimensional layer (layer II) with a separate distance $d$. A schematic of the structure is shown in Fig.~\ref{fig:schem}, where layer II can be a chiral (another graphene) layer or a nonchiral (2DEG) layer. We assume each layer is of zero thickness in the direction normal to the plane of the system at zero temperature. The layers are separated by a dielectric material (shown in Fig.~\ref{fig:schem}) with a dielectric constant $\epsilon_2$ and we suppose that the tunneling of electrons between the layers is negligible; however, the Coulomb interlayer interaction plays a role in the system.
The Hamiltonian of such a system can be written as~\cite{Profumo DLG}
\begin{eqnarray}\label{eq:Hamil}
\hat{H}=\hbar v_{\rm F} \sum_{k,\gamma,\beta} \hat{\psi}^\dagger _{k,\gamma}(\sigma_{\gamma\beta}.k)\hat{\psi}_{k,\beta}+\hat{T}_2 \nonumber \\
+\frac{1}{2S}\sum_{q\neq 0,l,l'}V_{ll'}(q)\hat{\rho}_{l}(q)\hat{\rho}_{l'}(-q)
\end{eqnarray}
where $\gamma, \beta$ are pseudospin indexes in the $x$ and $y$ directions, $l=1(2)$ is layer Index, $v_{\rm F}\simeq 10^6$m/s is the Fermi velocity of graphene and $\hat{\rho}_{l}(q)$ denotes the density operator in layer $l$ with the momenta $q$. $\hat{T}_2$ denotes the kinetic energy of layer II which is the same as the Dirac equation for a DLG system, while it is $\psi^\dagger_{k,2}\hbar^2\hat{k}^2/2m^*\psi_{k,2}$ in the case of a G-2DEG hybrid~\cite{Principi Hybrid} where $m^*$ is the electron band mass. The last term $V_{ll'}(q)$, refers to the inter- and intralayer electron-electron Coulomb interactions that can be obtained by using electrostatic relations for a two parallel conducting systems~\cite{Profumo DLG}. The intralayer interaction is given by
\begin{eqnarray}
V_{11}(q)=\frac{4\pi e^2}{q D(q)}[(\epsilon_2 +\epsilon_3)e^{qd}+(\epsilon_2 -\epsilon_3)e^{-qd}]
\end{eqnarray}
where
\begin{eqnarray}
D(q)=[(\epsilon_1 +\epsilon_2)(\epsilon_2 +\epsilon_3)e^{qd}+(\epsilon_1 -\epsilon_2)(\epsilon_2 -\epsilon_3)e^{-qd}]\nonumber
\end{eqnarray}
and the interlayer interaction is defined as
\begin{eqnarray}
V_{12}(q)=V_{21}(q) =\frac{8\pi e^2}{q D(q)} \epsilon_2
\end{eqnarray}
where $\epsilon_l$ denotes the dielectric constant of materials as illustrated in Fig.~\ref{fig:schem}. Furthermore, the interaction in layer II can be obtained by replacing $\epsilon_1 \leftrightarrow \epsilon_3$ in $V_{11}(q)$. Notice that, we are interested in the transport properties of layer I in the presence of layer II.

\subsection{Conductivity and Hall coefficient in the tilted magnetic field}

Applying a tilted magnetic field with a small deviation from the layers' plane $\theta << 1$, where $\theta$ is the angle between the plane and the applied magnetic field, has two impacts. First of all, the parallel component of the magnetic field, $B_{||}$ gives rise to the spin polarization of carriers in the system and changes the chemical potential by $\pm\Delta/2$ where $\Delta=g^*\mu_B B_{||}$ with $g^*$ being the effective Lande' $g$-factor and $\mu_B$ is the Bohr magneton. An enhanced $g$-factor, $g^*=2.7\pm 0.2$ for a single and bilayer graphene has been measured~\cite{kurgamov}. This enhancement is due to the impact of the electron-electron interaction in electronic liquid systems~\cite{asgari}. Notice that for $B_{||}=0$, the spin degeneracy is $g_s=2$ whereas for large $B_{||}$, the degeneracy factor is given by $g_s=1$ and for the intermediate fields, the system is partially spin polarized. Furthermore, the electron density for spin up and spin down are given by
\begin{eqnarray} \label{eq:npm}
n^\pm = \frac{g_\nu}{4\pi \gamma^2} (\mu \pm \Delta/2)^2
\end{eqnarray}
where $\mu$ is the chemical potential, $g_\nu$ is the valley degeneracy and $\gamma$ is $\hbar v_{\rm F}$.
By increasing the magnetic field, the number of electrons with spin up increases and therefore the number of electrons with spin down decreases up to the saturation field, $B_{S1}$ in which $n^-=0$. At this field $\mu=\Delta /2$, thus the saturation field is $B_{S1}=\sqrt{2}E_F/g^*\mu_B$ in which we use $n_1=n^{+}+n^- =E_{\rm F}^2/\pi\gamma^2$ where $E_{\rm F}$ is the Fermi energy.
Therefore, charge densities in term of the magnetic field can be changed and we thus have
\begin{eqnarray}
n^\lambda= \left \{ \begin{array}{rcl} \frac{n_1}{2}(\sqrt{1-B'^2/2} +\lambda \sqrt{2}/2 B')^2 & \mbox{if} & B'<1 \\
n_1(\delta_{+,\lambda}+\frac{1}{4B'^2}(B'^2-1)^2) & \mbox{if} & B'>1 \end{array}\right.
\end{eqnarray}
where $\lambda=\pm$, $\delta_{ij}$ is a Kronecker delta and $B'=B_{||}/B_{S1}$. Accordingly, $B_{||}$ leads to a decomposition of the conductivity into two different spin dependent channels, namely $\sigma_+$ and $\sigma_-$  and the total conductivity is $\sigma=\sigma_{+}+\sigma_-$ since the contributions of the spin up and down channels are parallel. Notice that in our study here, there is no mechanism to change spin direction.

It is worth mentioning that the magnetic field also affects the transport properties of layer II and leads to a change in the electron density distribution. For a DLG system, the aforementioned formula remains the same for layer II with the electron density $n_2$, since generally, the two layers can have different electron densities. If we consider a 2DEG system as layer II, the magnetic field polarizes the spin of the conduction electron up to a saturation field which is $B_{S2}=2\varepsilon_{\rm F}/g^*\mu_B$ where $\varepsilon_{\rm F}$ is the Fermi energy of the 2DEG layer. Increasing the magnetic field can not excite electrons from the valence band owing to the existence of a large band gap in the dispersion relation of the 2DEG system. Therefore, the electron density in such a layer is
\begin{eqnarray}
n^\lambda =\left \{ \begin{array}{rcl} \frac{n_2}{2} (1+\lambda B_{||}/ B_{S2} ) & \mbox{for}  & B_{||}<B_{S2} \\
n_2~\delta_{+,\lambda} & \mbox{for} & B_{||}>B_{S2} \end{array} \right.
\end{eqnarray}

The second consequence of the tilted magnetic field refers to the perpendicular component in which $B_\perp=B\sin(\theta)$. We assume that the field of strength is not strong enough to change the dispersion relation into the Landau levels producing the quantum Hall conductivity~\cite{MT} and therefore it only results in an ordinary Hall conductivity, $\sigma_{xy}$. Thus the electromagnetic fields, by using semiclassical relations~\cite{ziman}, are given by

\begin{eqnarray}
{\bf E}=\frac{1}{\sigma_+}( {\bf J}_++\beta_+ {\bf B}\times {\bf J}_+)
=\frac{1}{\sigma_-}( {\bf J}_-+\beta_- {\bf B}\times {\bf J}_-)
\end{eqnarray}
where $\beta_i=e\tau_i v_F / \hbar k_{{\rm F}}^i c$, $i=\pm$ indicates spin up or spin down and $k_{{\rm F}}^i$ denotes the Fermi wave vector of $i$th spin component with the electron density $n^{\pm}$.  by using a current vector such as ${\bf J_i}=a_1^i {\bf E}+a_2^i{\bf B} \times {\bf E}$ and the Eq. (7), it is easy to find $a_1^i$ and $a_2^i$. We thus have ${\bf J}=a{\bf E}-b{\bf B}\times {\bf E}$ in which ${\bf J}={\bf J}_++{\bf J}_-$ and
\begin{equation}\label{Rxy}
a=\sum_{i=+,-} \frac{\sigma_{i}}{1+\beta_i^2 B_\perp ^2}\,\,\, , b=\sum_{i=+,-} \frac{\sigma_{i} \beta_i}{1+\beta_i^2 B_\perp ^2}
\end{equation}
Now, by considering the electric field, ${\bf E}=R_{xx} {\bf J} + {R_H} {\bf B}\times {\bf J}$, we calculate the magnetoresistance and Hall coefficient expressions as
\begin{eqnarray}\label{Rxy}
R_{xx}&=&\frac{a}{a^2+b^2B_\perp ^2}\nonumber\\
R_H&=&\frac{b}{a^2+b^2 B_\perp ^2}\end{eqnarray}

$R_{xx}$ is the longitudinal resistivity and ${R_H}=R_{xy}/B_\perp$ is the Hall coefficient which depend on the conductivities of spin up and spin down which will be explored in the next section within the Boltzmann approximation.

Before describing the Boltzmann equation, it is worth mentioning that the conductivity of a two-component system can be written as $\sigma = \begin{pmatrix} \sigma_1 & \sigma_D \\ \sigma_D & \sigma_2 \end{pmatrix}$
where $\sigma_D$ is the drag conductivity.
To calculate the resistivity, one should take the inverse of the conductivity matrix
\begin{eqnarray}
\rho = \begin{pmatrix} \rho_1 & \rho_D \\ \rho_D & \rho_2 \end{pmatrix} =
\frac{1}{\sigma_1 \sigma_2 -\sigma_D^2} \begin{pmatrix} \sigma_2 & -\sigma_D \\ -\sigma_D & \sigma_1 \end{pmatrix}
\end{eqnarray}
Using the fact that $\sigma_D^2 << \sigma_1 \sigma_2$, the resistivity of each layer is obtained as $\rho_i=1/\sigma_i$ and we therefore define the inverse of the conductivity as the resistivity in both longitudinal and Hall conductivities of layer I.

\subsection{Boltzmann approach for the conductivity}
To calculate the conductivity of layer I in which the electron density, $n_1$ is much larger than the concentration of impurities $n_{imp}$, we therefore can use the Boltzmann approach which is based on the relaxation time approximation. Our theory is formulated for weak disorder and we consider only charged impurity scattering without spin-flip processes. In this approach, the conductivity of the system is given by~\cite{Dassarmarev}
\begin{equation} \label{eq:sigma}
\sigma_\pm =\frac{e^2}{2}\hbar ^2 v_F^2  \int d\varepsilon D^{\pm}(\varepsilon) \tau(\varepsilon) [-\frac{\partial n_F(\varepsilon)}{\partial \varepsilon}]
\end{equation}
where $n_F(\varepsilon)$ is the Fermi distribution function. At zero temperature, the above equation reduces~\cite{ref:dassarma1,Dassarmarev} to $ \sigma_\pm = \frac{e^2 v_F^2 D(E_{\rm F}^\pm) \tau (E_{\rm F}^\pm)}{2}$
, where $D(E_{\rm F}^\pm)$ is the density of states at the Fermi energy and in the case of graphene $D(E_F^\pm)=g_sg_\nu E_F^\pm / 2\pi \gamma^2$.
$\tau (E_k)$ is the relaxation time which is, for graphene layer, given by~\cite{ref:dassarma11}
\begin{eqnarray}\label{eq:tau}
\frac{1}{\tau(E_k)}=\frac{\pi}{\hbar}\sum_{k'} n_{imp}|W_{11}(q)|^2 (1-\cos ^2(\theta)) \delta(E_k - E_{k'})
\end{eqnarray}
where $q=|k-k'|$, $\theta = \theta_k-\theta_{k'}$ and $W_{11}(q)$ represents the effective interaction between impurities and electrons. Short-range disorder plays a role in resistive scattering at higher carrier density and also high mobility~\cite{Dassarmarev}, however for mid-range densities, the main scattering source is the charged impurity disorder~\cite{ref:Ando, Alekseev}. We thus neglect all disorder mechanisms other than random charged impurities in the system and assume the charged impurities to be of random negative sign of units of the electron charge.
We assume that the charged disorder with density $n_{imp}$ is located in the plane of layer I.
Changing the variable $\sin(\theta/2)=y$ in Eq.~(\ref{eq:tau}) and writing the relation for different doping $n^\pm$ separately, the relaxation time reads
\begin{eqnarray}\label{eq:tauG}
\frac{1}{\tau(E_{\rm F}^\pm)}=\frac{4 n_{imp}}{\pi \hbar}\frac{E_F^\pm}{\gamma ^2} \int _0 ^1 dy |W_{11}(2k_F^\pm y)|^2 y^2\sqrt{(1-y^2)}\nonumber\\
\end{eqnarray}
where the relaxation time is evaluated at the $E_{\rm F}^\pm$ of layer I with the electron density $n^\pm$ and the Fermi wave vector $k_{\rm F}^\pm$. We use a minimal theory to describe effects of the parallel magnetic field.
The effective interaction in the studied layer takes the form~\cite{Profumo DLG}
\begin{eqnarray}
W_{11}(q)=\frac{V_{11}(q)+[V_{12}^2(q)-V_{11}(q)V_{22}(q)]\Pi^0_2(q)}{\varepsilon(q)}
\end{eqnarray}
where the dielectric function $\varepsilon (q)$ within RPA is
\begin{eqnarray}
\varepsilon(q)=[1-V_{11}(q)\Pi^0_1(q)][1-V_{22}(q)\Pi^0_2(q)]
\nonumber\\
-V^2_{12}(q)\Pi^0_1(q)\Pi^0_2(q)
\end{eqnarray}
where $\Pi^0_l(q)$ is the static charge response function of the $l$th layer and it is $D_F^+ P(q/k_F^+)+D_F^- P(q/k_F^-)$ for a doped graphene in the spin-polarized case, where
\begin{eqnarray}
P(y)= -\left \{ \begin{array}{rcl} &&~ 1~~~~ \mbox{if}~  y\leq2 \\
&&1+\frac{\pi y}{8}-\frac{1}{2}\sqrt{1-\frac{4}{y^2}}-\frac{y}{4} \sin^{-1}(\frac{2}{y})~~\mbox{if}~ y>2 \end{array}\right.
\end{eqnarray}

In the case that layer II is a 2DEG layer, we then need to modify the dielectric function by using $\Pi_2^0 (q) = \Pi_2^+ (q) + \Pi_2^- (q)$, where
\begin{eqnarray}
\Pi_2^\pm (q) = D [1-\sqrt{1-(2k_{\rm F2}^{\pm} /q)^2}\theta (q-2k_{\rm F2} ^\pm)]
\end{eqnarray}
the density of states is $D=g_v m^*/2\pi \hbar ^2$ and the Fermi wave vector is $k_{\rm F2}^{\pm}=\sqrt{4\pi n_2^{\pm}/g_s}$ .

By using Eq.~(\ref{eq:tauG}), we can write the magnetoresistance of layer I in the presence of another layer as
\begin{eqnarray}\label{eq:I}
\frac{\sigma}{\sigma^0}=\frac{1}{2}(\frac{I^0}{I^+}+\frac{I^0}{I^-})
\end{eqnarray}
where $I^\pm=\frac{\sqrt{\pi(n_1+n_2)}}{2\pi e^2} \int_0^1 dy y^2 \sqrt{1-y^2} \,W_{11}^2(2k_F^\pm y)$. Notice that  $\sigma^0$ and $I^0$ are the conductivity and $I^+$ of layer I when $B=0$.

\section{Numerical result and discussion}\label{sec3}

\begin{figure}
\includegraphics[width=1.0\linewidth]{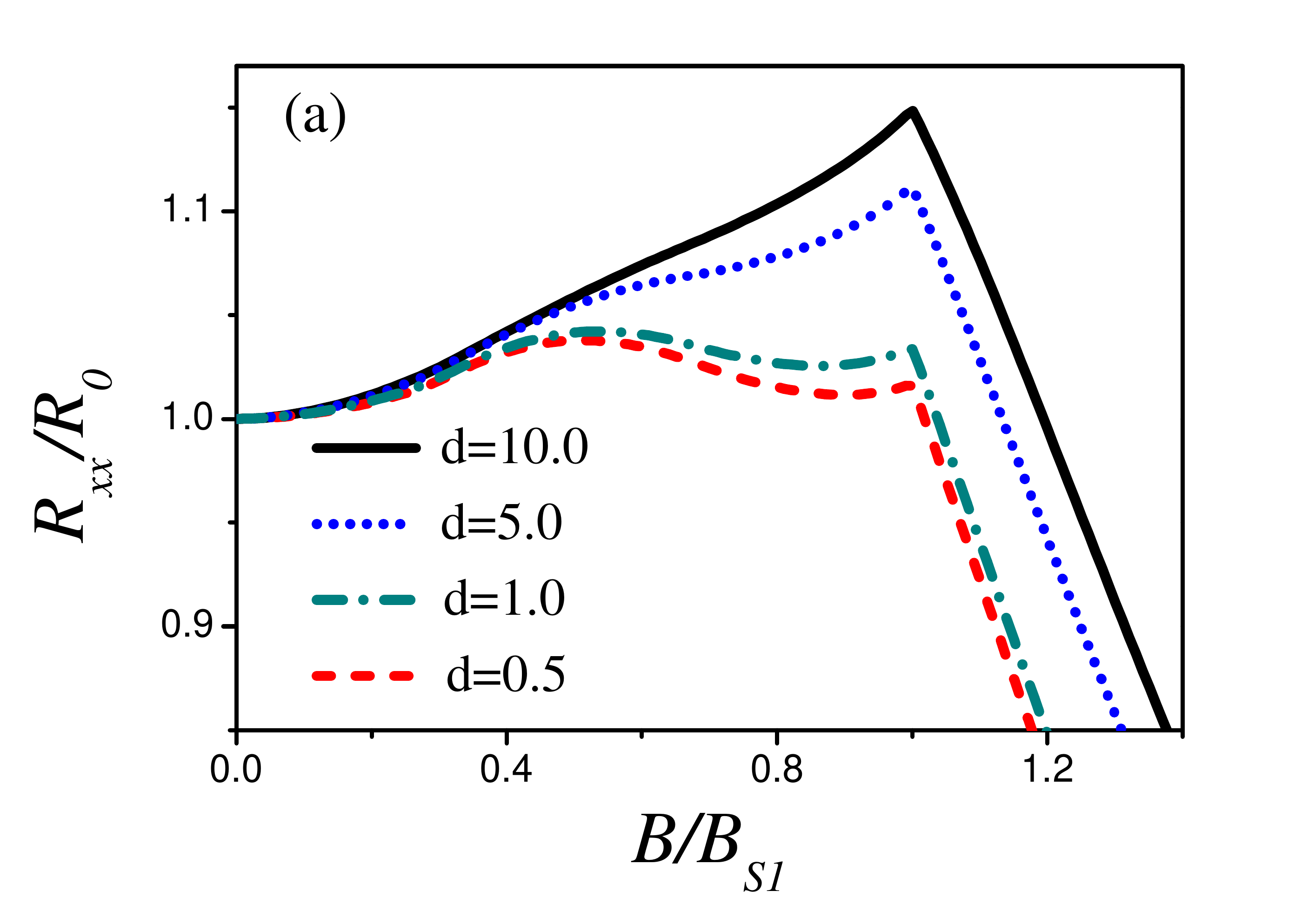}
\includegraphics[width=1.0\linewidth]{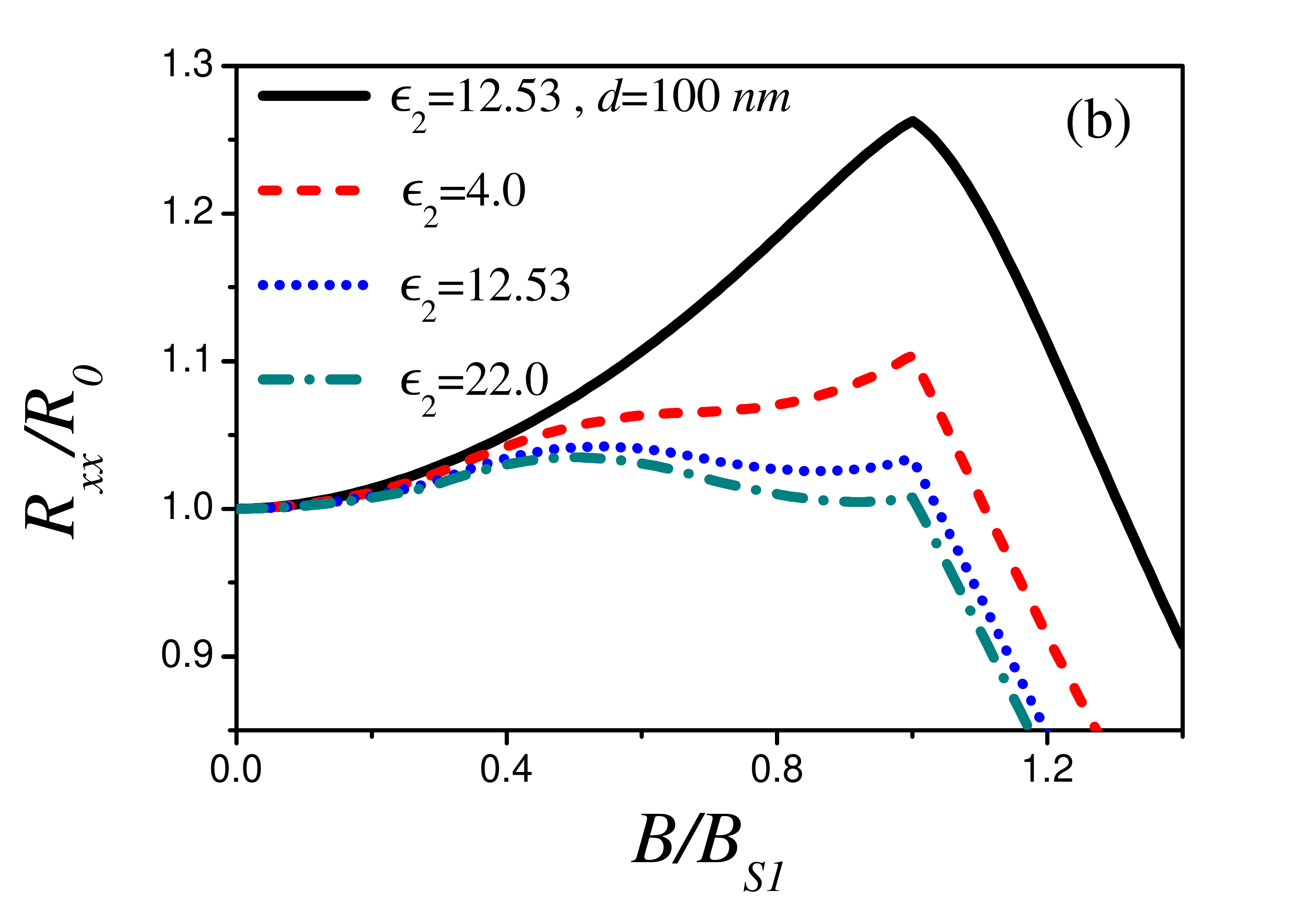}
\caption{(Color online) Longitudinal resistivity of layer I in a hybrid double-layer system composed by the doped graphene sheet that is Coulomb coupled to another doped graphene in which they are separated by a dielectric material as a function of $B=B_{||}$ for (a) different distance between layers in units of nm for $\epsilon_2=12.53$ and (b) different value of $\epsilon_2$ with constant distance $d=1.0$nm. We also consider a plot with $d=100$nm to suppress the interlayer effect. We choose $\epsilon_1=1.0$, $\epsilon_3=12.53$ and the electron densities are $n_1=10^{11}$cm$^{-2}$ and $n_2=10^{10}$cm$^{-2}$, respectively.
\label{fig:RxxB}}
\end{figure}

\begin{figure}
\includegraphics[width=1.0\linewidth]{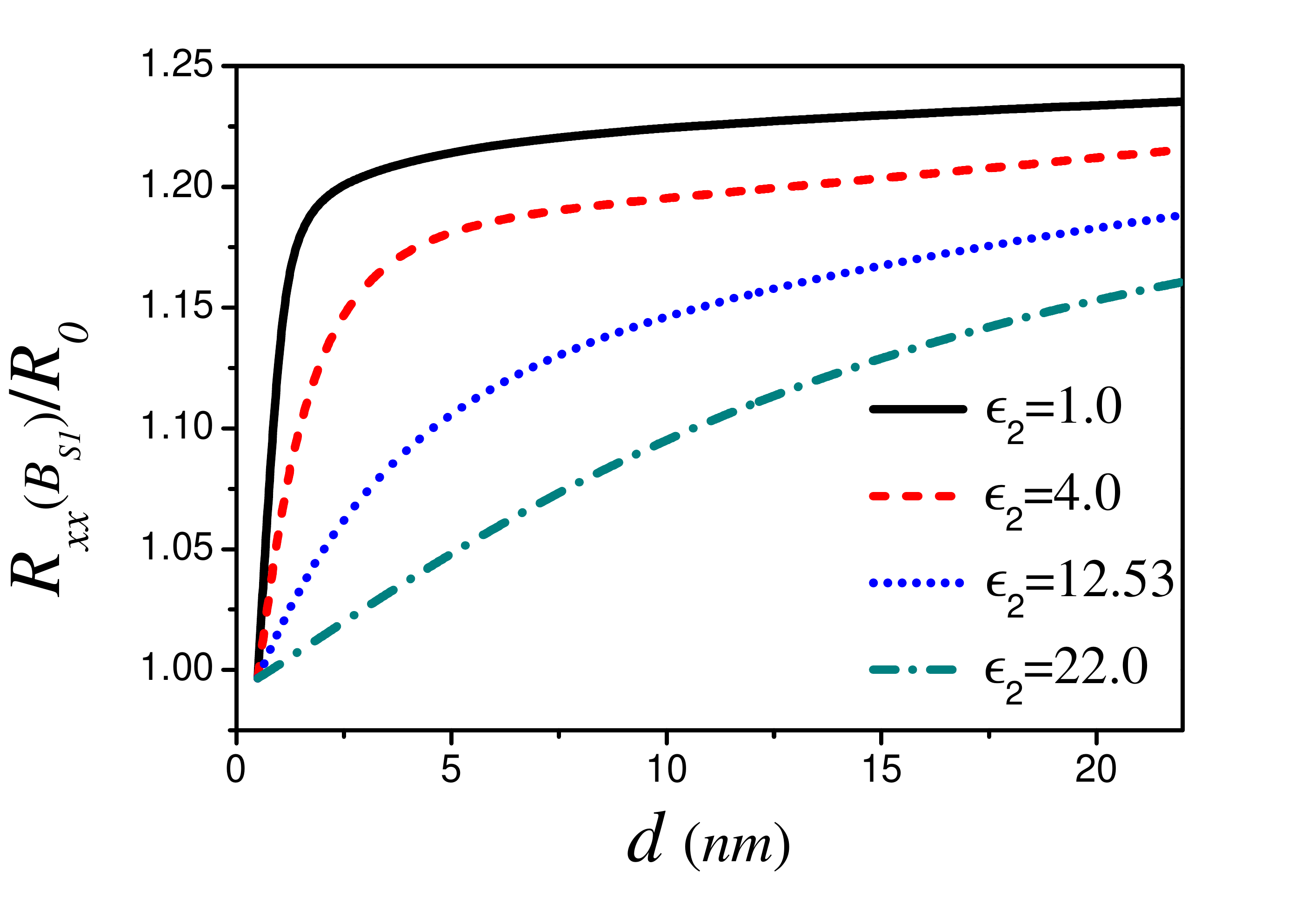}
\caption{Longitudinal resistivity of layer I at the saturation field versus the distance between layers in units of nm in a hybrid double-layer graphene system for different values of $\epsilon_2$. We choose $\epsilon_1=1.0$, $\epsilon_3=12.53$ and the electron densities are $n_1=10^{11}$cm$^{-2}$ and $n_2=10^{10}$cm$^{-2}$, respectively.
\label{fig:Rxxd}}
\end{figure}

In this section, we present our main numerical results based on the theory presented in the previous section. Our aim is to explore the impact of the correlation effects on the magnetoresistance and Hall coefficient of layer I. Layer II can be either a gapless graphene or a 2DEG layer. We therefore study the structures by considering both parallel and tilted magnetic fields. In all numerical results we use $e^2/\gamma=2.2$ and $g^*=2$.

\subsection{Double-layer graphene in a parallel magnetic field, $B_{\bot}=0$}

We consider a hybrid double-layer graphene system (DLG) in which the layers are coupled within the interlayer Coulomb interaction and they are separated by a dielectric material, as schematically illustrated in Fig.~1. We show the resistivity of layer I as a function of the $B=B_{||}$ in Fig.~\ref{fig:RxxB} where electron densities are $n_1=10^{11}$ and $n_2=10^{10}$ cm$^{-2}$, respectively. The resistivity is scaled by its value at zero magnetic field denoted by $R_0$ and it raises by increasing the magnetic field owing to the suppression of the screening and increases of the effective interaction screened within the RPA dielectric function. As $B$ increases beyond its saturation field value, $B_{S1}=\hbar v_{\rm F} \sqrt{2\pi n_1}/(g^*\mu_B)$ in which layer I is fully spin polarized, the hole density is created in the valence band and makes $R_{xx}$ decrease sharply, which demonstrates the negative differential magnetoresistance. We analyze the resistivity for the different layer distances, $d$ and the media dielectric constants, $\epsilon_2$ and our numerical results are shown in Fig~\ref{fig:RxxB}(a) and 2(b) respectively. Interestingly enough, Fig.~\ref{fig:RxxB}(a) shows that the resistivity of layer I is significantly modified when $n_2<n_1$. For $B>B_{S2}$, the effective interaction decreases which gives rise to a decrease of the resistivity. Our numerical results show that the resistivity has a negative differential magnetoresistance for a small $d$ value and for a region in which $B<<B_{S1}$. This region depends on the electron density in layer II, $n_2$ and the slop of the $R_{xx}$ changes by decreasing $n_2$ at small $B$. Thus, a larger conductivity of layer I can be reached by using $n_2<n_1$. It is obvious that we also get the resistivity of a single graphene layer by considering a very large $d$ value.

Moreover, it has been shown that~\cite{Hosono} the dielectric constant between the two layers in double-layer graphene systems has a significant role in transport properties. Although increasing the dielectric constant screens the interlayer interaction, it has a significant role on the intralayer screening of the considered layer too. Accordingly, a competition between the effect of the inter- and intralayer interactions on the magnetoresistance of layer I is vital when dielectric materials with higher $\epsilon_2$ are used. We also obtain the reduction of the resistivity when the dielectric constant increases. We choose three different materials, namely h-BN, Al$_2$O$_3$ and HfO$_2$ for $d=1$nm to explore the effect of the interlayer interaction on the magnetoresistance of layer I in the DLG system. In the same manner, we obtain the negative differential magnetoresistance in area which $B<B_{S1}$ due to the correlation effects induced in layer I from layer II.

We analyze the resistivity at the saturation magnetic field for layer I as a function of $d$ in Fig.~\ref{fig:Rxxd}. We obtain a resistivity enhanced by about $20$\% by increasing the layer distance for the case in which there is a larger dielectric constant between layers owing to the long-ranged interlayer interaction and the fact that $V_{12}(q)$ is proportional to $\epsilon_2$. The $R_{xx}(B_{S1})$ significantly changes for a small $d$ and $\epsilon_2$.

\subsection{Graphene-2DEG hybrid in a parallel magnetic field, $B_{\bot}=0$}

\begin{figure}
\includegraphics[width=1.0\linewidth]{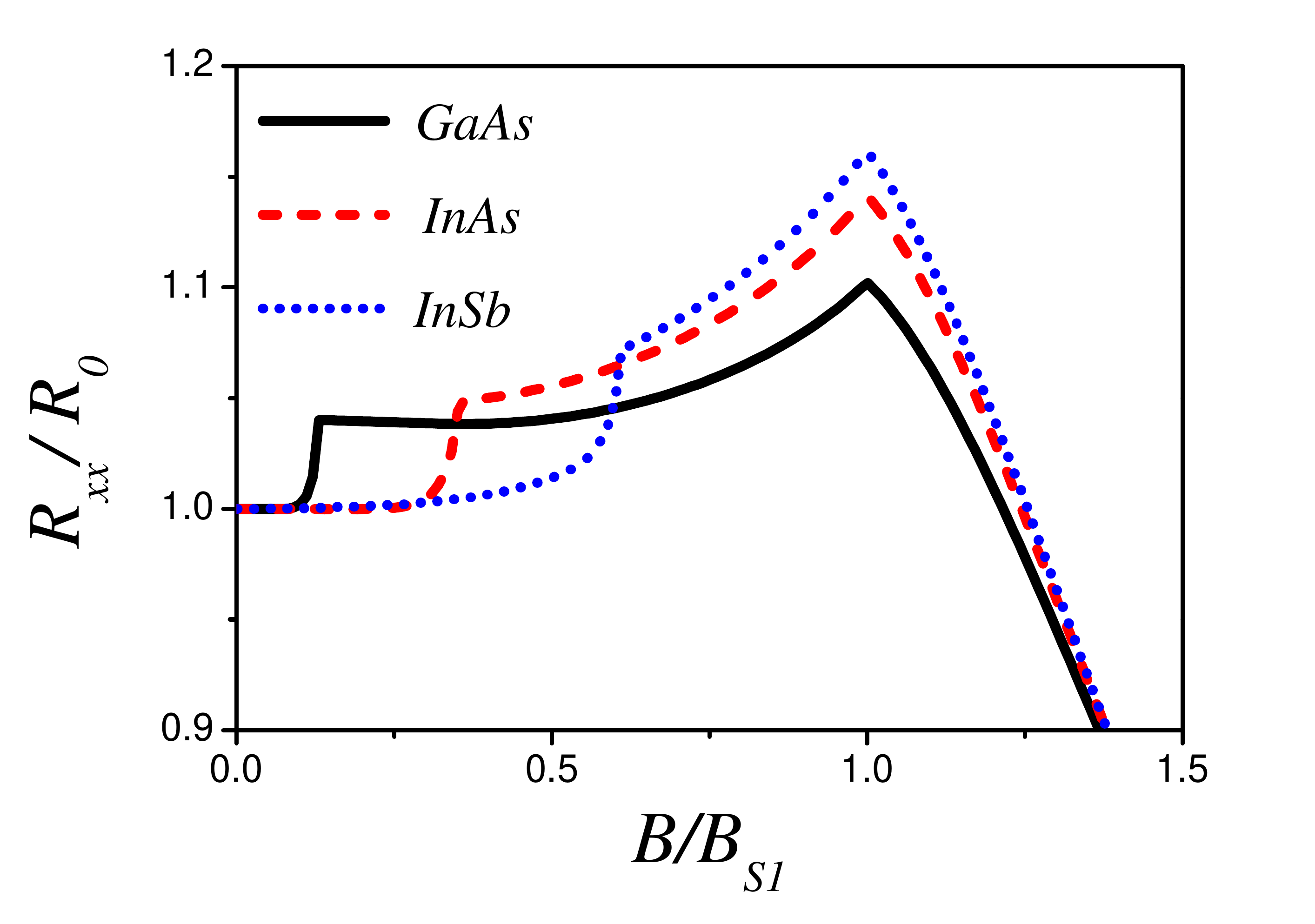}
\caption{(Color online) Magnetoresistance of layer I as a function of the $B=B_{||}$ in the presence of a 2DEG layer for different 2DEG materials.
Here, the dielectric constants of 2DEG systems are chosen as~\cite{ashcroft}: $10.9, 12.2$ and $15.7$ in addition, the effective mass of these materials are the electron band effective mass which is $0.07m$, $0.026m$ and $0.015m$ where $m$ is the electron bare mass for GaAs, InAs and InSb, respectively. The electron density of the both layers is the same and equal to $10^{11}$ cm$^{-2}$, $\epsilon_1=1$, $\epsilon_2=\epsilon_3$, which are equal to the dielectric constant of the 2DEG layer and the distance between layers is $d=10$nm.
\label{fig:RxxB2}}
\end{figure}

\begin{figure}
\includegraphics[width=1.0\linewidth]{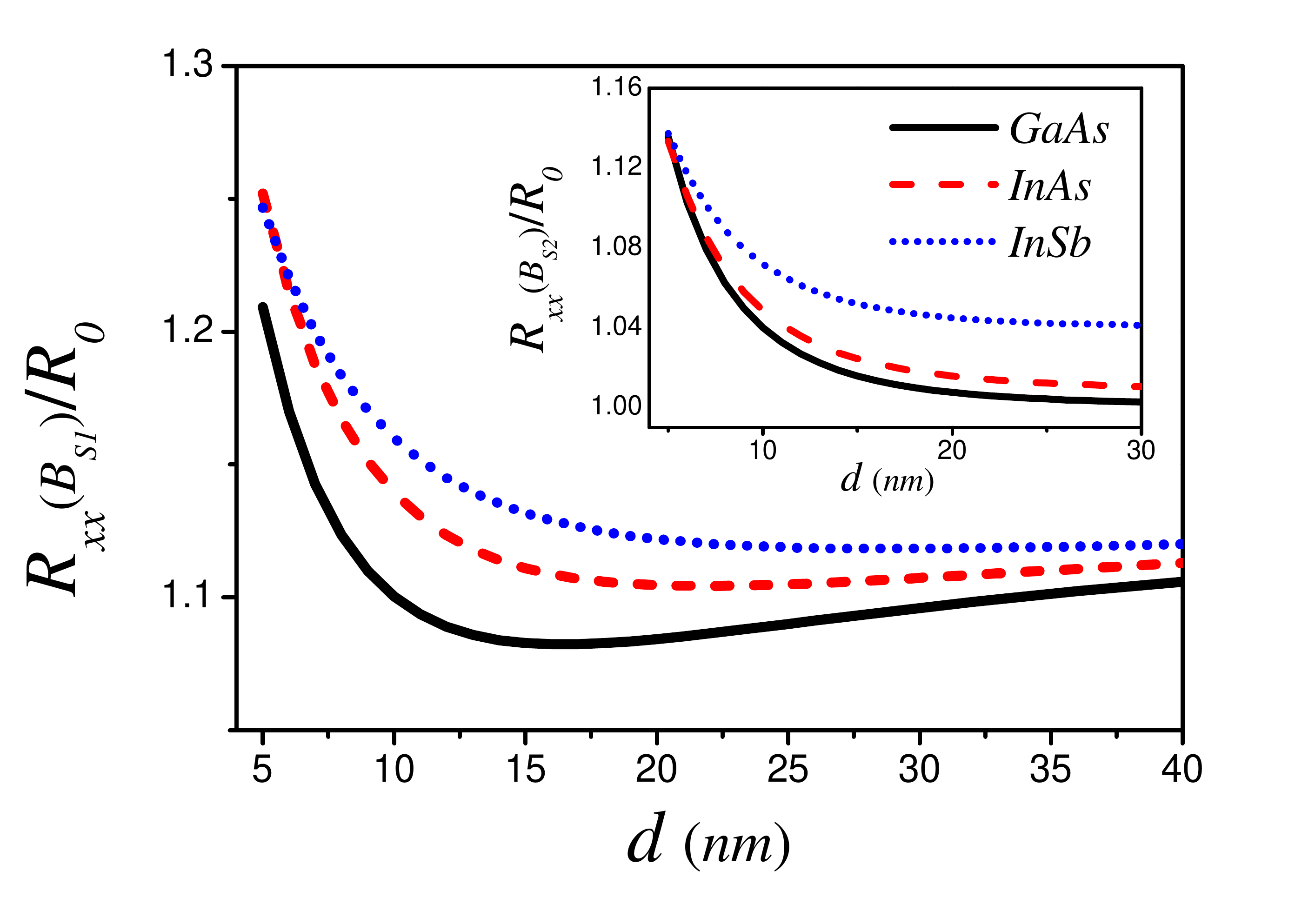}
\caption{(Color online) Value of the saturation field resistivity of layer I scaled by its zero-field resistivity as a function of $d$ in the presence of different 2DEG materials. Inset is the same for the saturation field of layer II. All other parameters are the same as Figure~\ref{fig:RxxB2}.
\label{fig:Rxxds1}}
\end{figure}

In this section, we consider a G-2DEG system in which a dielectric material is filled in the space between the two layers. For such a system, the saturation magnetic field of the 2DEG layer, $B_{S2}=\frac{2\pi \hbar ^2 n_{2}}{g_v m^* g^*\mu_B}$ is smaller than the corresponding one in a graphene layer system, $B_{S1}$ even though $n_1=n_2$.

Figure~\ref{fig:RxxB2} shows the longitudinal magnetoresistance of layer I in the presence of the 2DEG layer for different materials~\cite{ashcroft}. The electron density of both layers is the same and equal to $10^{11}$ cm$^{-2}$, $\epsilon_1=1$ and $\epsilon_2=\epsilon_3$ for $d=10$nm. This figure shows that because 2DEG carriers are fully polarized, the magnetoresistance of layer I exhibits a sharp jump at $B_{S2}$ which occurs at a much smaller value than the $B_{S1}$ value. The longitudinal resistivity increases by increasing $\epsilon_2$ due to the increased screening. Importantly, this feature differs from that in a DLG structure. Meanwhile, since $B_{S2}\sim m^{*-1}$, the position of the first peak of the $R_{xx}$ decreases by increasing the electron effective mass.

In order to bring out the difference between G-2DEG and DLG systems, we plot the magnetoresistance of layer I at the saturation magnetic field, $B_{S1}$ in Fig.~\ref{fig:Rxxds1} when the 2DEG layer is placed near the graphene layer. The results show that the $R_{xx}$ decreases at small $d$, behaves nonmonotonically as a function of $d$ and the results are in different trend as compared with Fig.~3 in which the results for DLG systems are shown. The difference can be understood in terms of the effect of the chirality effect in the charge-charge response function of the doped graphene~\cite{barlas}.

\subsection{Longitudinal resistivity in a tilted magnetic field, $B=\sqrt{B^2_{||}+B^2_{\bot}}$}

\begin{figure}
\includegraphics[width=1.0\linewidth]{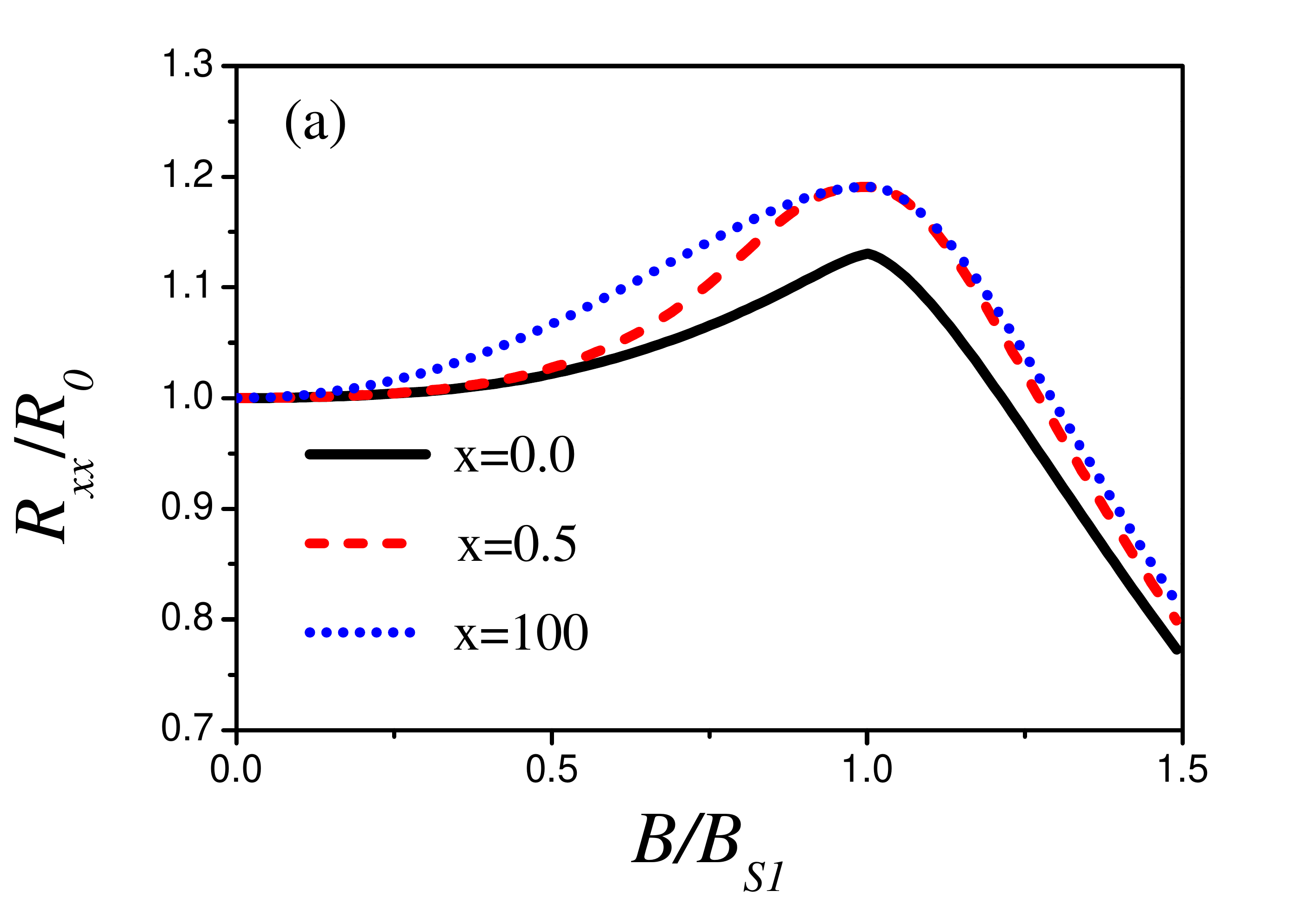}
\includegraphics[width=1.0\linewidth]{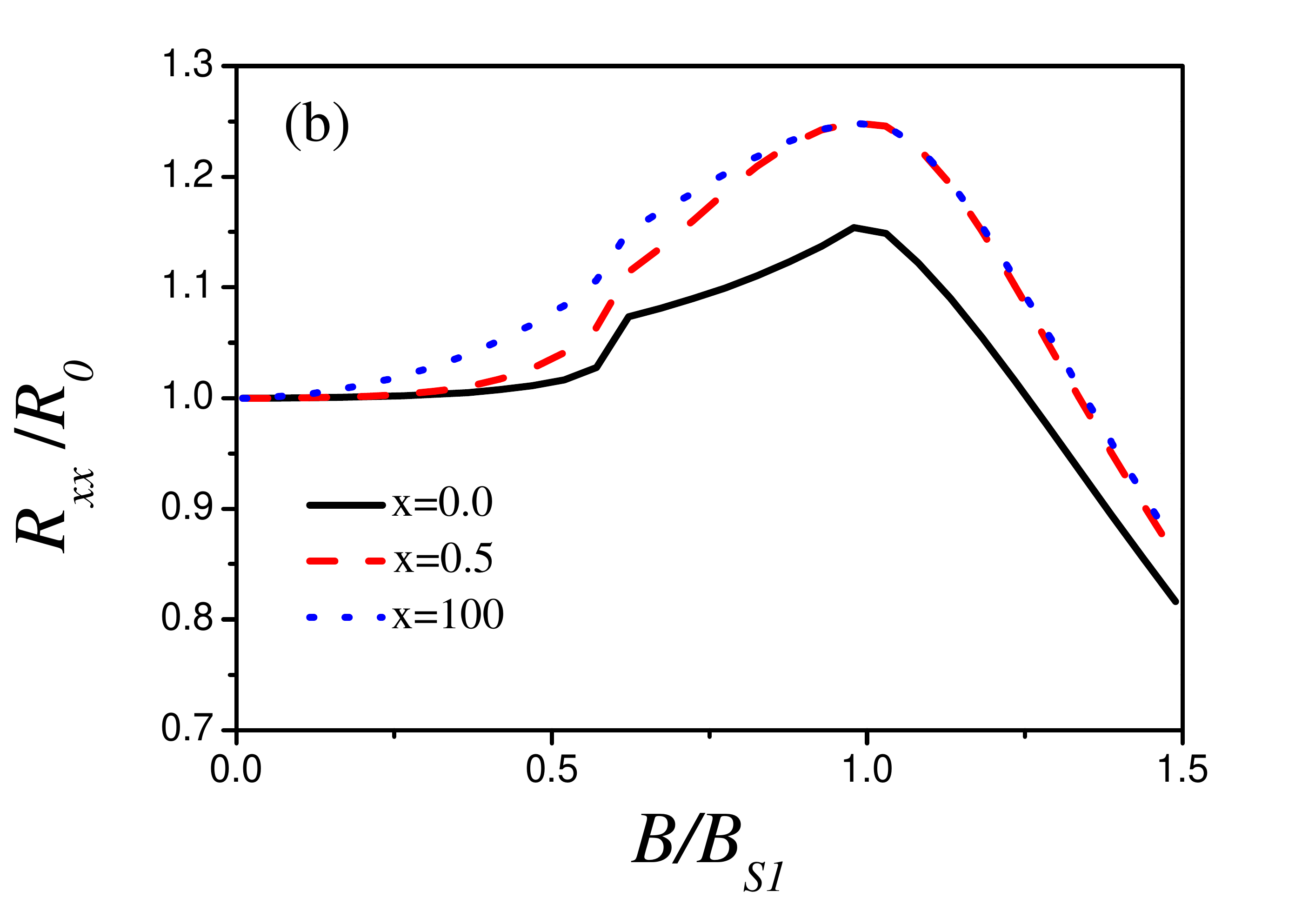}
\caption{(Color online) Longitudinal magnetoresistivity of a system consists of a single graphene layer ($\epsilon_2=12.52$, $\epsilon_1=1.0$) with $n_1=10^{12}$cm$^{-2}$(a) and in a hybrid system in which a G-2DEG system for InSb with $n_1=n_2=0.1\times 10^{12}$cm$^{-2}$, $\epsilon_3=15.7$ and $d=10$nm (b) as a function of the magnetic field amplitude, $B=\sqrt{B^2_{||}+B^2_{\bot}}$. The magnetoresistance is calculated for different values of the tilting parameter $x$.
\label{fig:Rxx}}
\end{figure}

In the presence of a tilted magnetic field, $B=\sqrt{B^2_{||}+B^2_{\bot}}$ where $B_{||}\gg B_{\bot}$, we should use the full expression of the conductivity given in Eq.~(\ref{Rxy}) and it can be written as
\begin{equation}
 B_\perp \beta_{\pm} = \frac{B I_0n_t}{2B_{S1}I^\pm~n^{\pm}_1}~x=\alpha_{\pm}~x
\end{equation}
where $I^\pm$ is the function given under Eq. (\ref{eq:I}), and $n_t$ is the total density of the two layers ($n_1+n_2$). Here, we define a new parameter~\cite{Herbut}, $x=0.895 \sin(\theta) \sqrt{\bar{n}_1}/\bar{n}_{imp} I_0$ ($\bar{n}_{1}$ and $\bar{n}_{imp}$ are in units of $10^{12}$cm$^{-2}$). Note that $x$ depends on both the angle as well as $n_{imp}$, therefore larger values of $x$ can be obtained by using clean samples.

In order to understand the impact of the tilted magnetic field on the resistivity, we first examine $R_{xx}$ of an isolated graphene sheet in the presence of $B$. Figure~\ref{fig:Rxx}(a) shows the effect of the tilted magnetic field on the longitudinal resistivity of an isolated $n-$doped graphene sheet. The longitudinal resistivity increases by increasing the value of $x$. The magnitude of differential magnetoresistance changes significantly in comparison with that result obtained when $B_{\bot}=0$. Moreover, the maximum of the $R_{xx}$ at the $B_{S1}$ is not sensitive to the finite value of $x$ and shows a good agreement with those trends obtained in experiment~\cite{kurgamov, Han}. It should be noted that the maximum of the $R_{xx}$ is a good quantity to explore quantum states in a graphene system and for this purpose, the magnetoresistance of the insulating state that forms at the charge neutrality point in the presence of tilted magnetic field has been measured~\cite{zhao} and data indicated that the zero-filling factor quantum Hall state in single layer graphene is not spin-polarized. We choose $x=100$ for an extremely clean sample and this figure shows that the $R_{xx}$ increases slightly by increasing large value of $x$. Having known the behavior of $R_{xx}$ as a function of $B$ for a single graphene sheet, we could explore $R_{xx}$ for layer I in a hybrid system. We consider only a G-2DEG system for an InSb layer with $n_1=n_2=0.1\times 10^{12}$cm$^{-2}$, $\epsilon_3=15.7$ and $d=10$nm and $R_{xx}$ is shown as a function of $B$ in Fig.~\ref{fig:Rxx}(b). As it is seen, $R_{xx}$ increases with increasing $x$ value and the fact that $B_{S2}<B_{S1}$, a sharp change in the results at $B_{S2}$ still remains. These numerical results exhibit that the interlayer interaction has a more significant effect specially for a small value of $x$.

\subsection{Hall resistivity in a tilted magnetic field}

\begin{figure}
\includegraphics[width=1.0\linewidth]{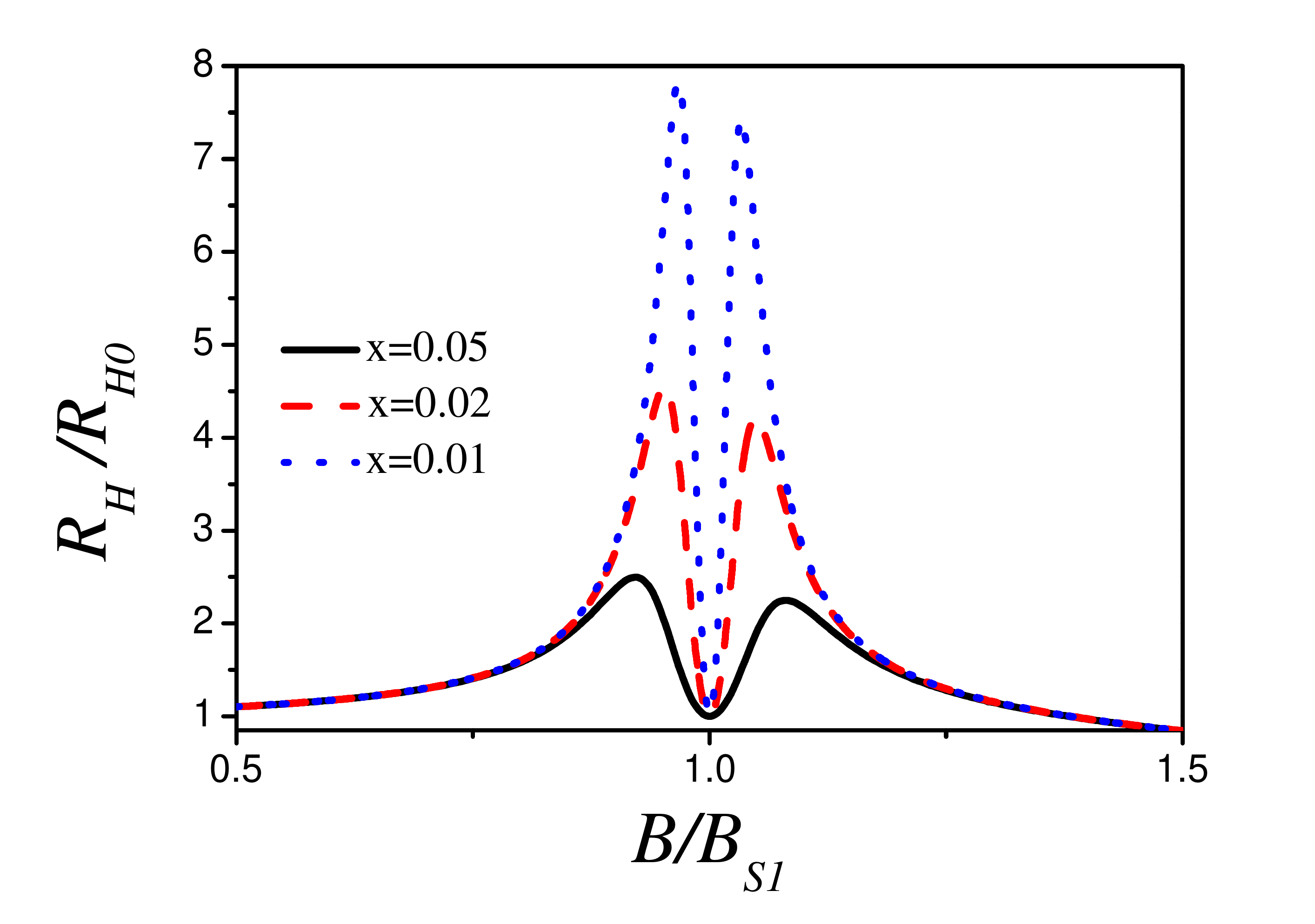}
\caption{(Color online) Hall coefficient of layer I as a function of the magnetic field amplitude, $B=\sqrt{B^2_{||}+B^2_{\bot}}$ in the presence of the 2DEG $InAs$ for different values of $x$ at given constant $\theta=4^0$. We consider $\epsilon_3=\epsilon_2=12.2$, $m^*=0.026 m_e$ and $n_1=0.1\times 10^{12}$, $n_2=0.01\times 10^{12}$cm$^{-2}$. The layers are separated with a distance $d=10$nm.
\label{fig:Rxy2}}
\end{figure}

Finally, by applying the tilted magnetic field, $B=\sqrt{B^2_{||}+B^2_{\bot}}$ in which there is a weak perpendicular component $B_\perp$ and using two-species conductivities $\sigma_+$ and $\sigma_-$, we can straightforwardly calculate the Hall coefficient $R_H=R_{xy}/B_\perp$ from Eq.~(\ref{Rxy}).
Normalizing the Hall coefficient to its zero magnetic field value, $R_{H0}=R_H({B\rightarrow 0})$ one gets
\begin{equation}
R_H/R_{H0}=\frac{b}{a^2+b^2B_\perp^2}\frac{\sigma_0}{\beta_0}
\end{equation}

Note that $\beta_+=\beta_-$ for the limit of $B_{||}\rightarrow 0$ and $\beta_0=\beta_+(B=0)$. The Hall coefficient of a 2DEG system shows a peak before the saturation magnetic field point and reaches to a constant value at the saturation magnetic field~\cite{Herbut}. Moreover, the Hall coefficient of the 2DEG system increases monotonically by decreasing the $x$ value, however our numerical results show that the Hall coefficient of layer I is significantly enhanced by decreasing the $x$ values.

In Fig.~\ref{fig:Rxy2}, the behavior of the Hall coefficient of layer I as a function of the $B=\sqrt{B^2_{||}+B^2_{\bot}}$ is shown in the presence of the InAs 2DEG for different values of $x$. Same qualitative results can be obtained for a single graphene layer and therefore we conclude that the Hall coefficient of graphene differs from that result of the 2DEG system. As is obvious in this figure, the two peaks occur at finite $x$. Importantly, if we interpret the Hall resistivity as an effective density of charge carriers, our numerical  results will show that the effective density of carriers changes more sharply around the $B_{S1}$. Nevertheless, the Hall coefficient in layer I has two peaks at finite $x$ values in which one peak occurs just below the $B_{S1}$ value and another takes place above that. These peaks become sharper at smaller $x$, accordingly a giant like Hall coefficient behavior of layer I can be reached at a small value of $x$. Notice that at $B\simeq B_{S1}$ the Hall coefficient decreases and eventually reaches to $R_{H0}$ at $B=B_{S1}$. To understand these features, we analytically calculate the $R_{H}$ in different regions. Let us start off by using the definition of the Hall coefficient of layer I

\begin{widetext}
\begin{eqnarray}\label{eq:1}
\frac{R_H}{R_{H0}}&=&\frac{\sigma_0}{\beta_0}\frac{b}{a^2+b^2 B_\perp^2}\nonumber\\
&=&\frac{D\sigma_0}{\beta_0}\frac{\sigma_+\beta_+(1+x^2\alpha_-^2)+\sigma_-\beta_-(1+x^2\alpha_+^2)}{(\sigma_+(1+x^2\alpha_-^2)+\sigma_-(1+x^2\alpha_+^2))^2+x^2(\sigma_+\alpha_+(1+x^2\alpha_-^2)+\sigma_-\alpha_-(1+x^2\alpha_+^2))^2}
\end{eqnarray}
\end{widetext}
where $D=(1+x^2\alpha_+^2)(1+x^2\alpha_-^2)$.
Notice that, in the above equation, there is a competition between $x$ and $\alpha_-$ since $\alpha_-$ is proportional to  $1/n^-$ and diverges at $B=B_{S1}$ where $n^-=0$, thus the zero limit of $x$ can not prevent this singular behavior. We take $x\alpha_-\rightarrow 0$ around $B_{S1}$. Notice that $\sigma_-$ has a finite nonzero value even at $B=B_{S1}$~\cite{Dassarmarev}.

\begin{figure}
\includegraphics[width=1.0\linewidth]{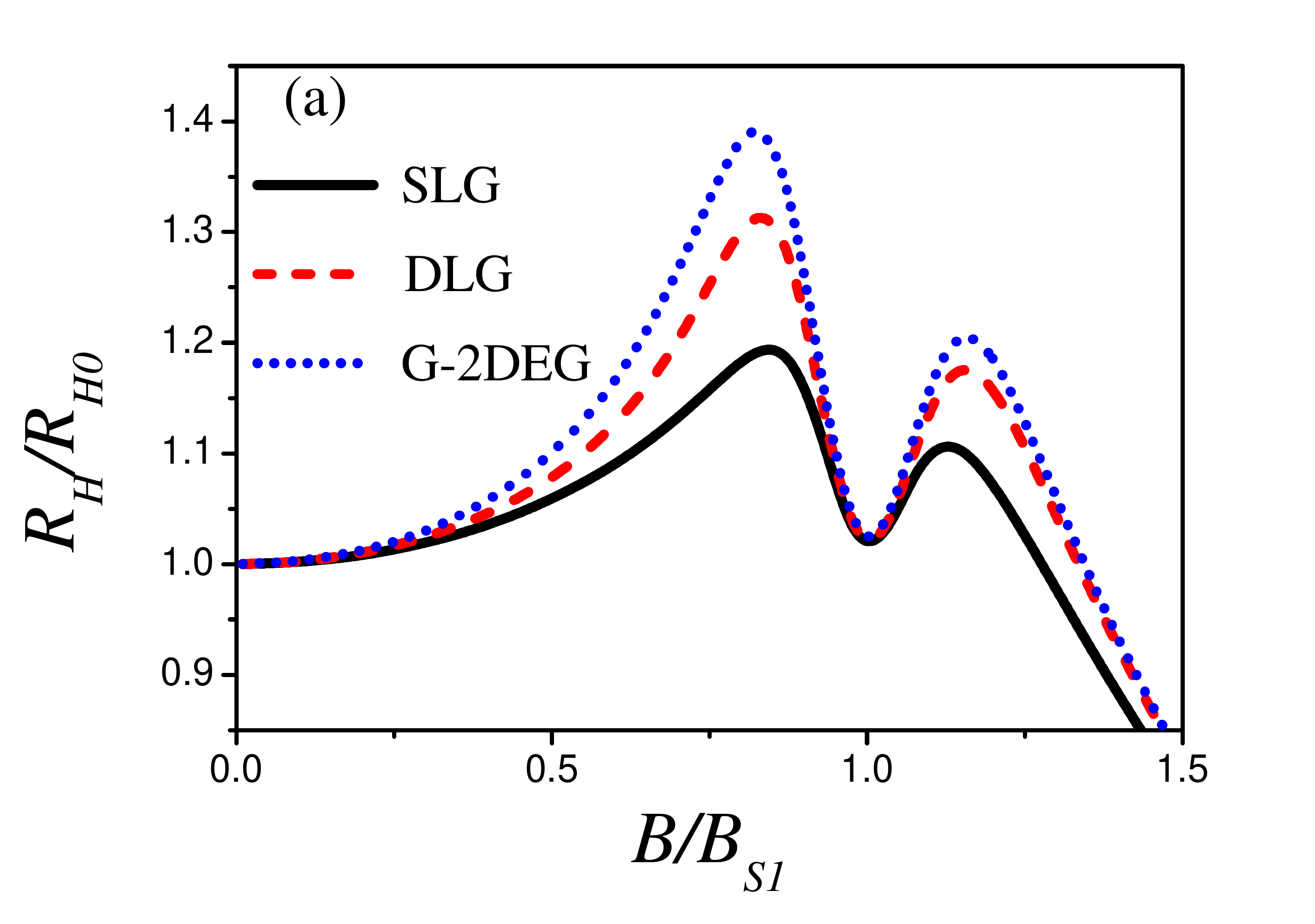}
\includegraphics[width=1.0\linewidth]{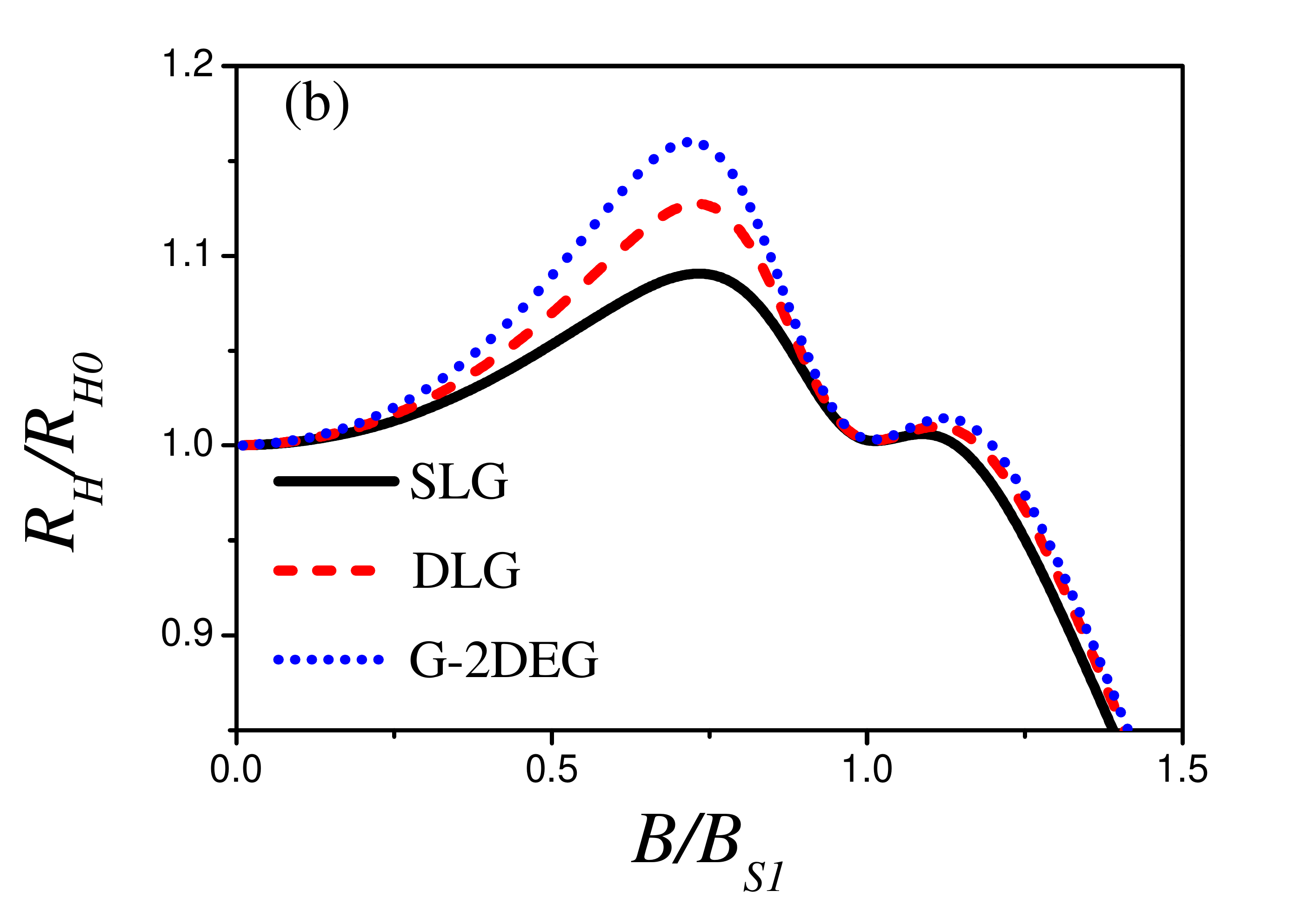}
\caption{(Color online) Hall coefficient of a single layer graphene (SLG) and Hall coefficient of layer I in a DLG and G-2DEG systems as a function of the magnetic field amplitude, $B=\sqrt{B^2_{||}+B^2_{\bot}}$ at given constant $\theta=4^0$ for (a) $x=0.2$ and (b) $x=0.5$. The electron densities are $n_1=0.1\times 10^{12}$, $n_2=0.01\times 10^{12}$cm$^{-2}$ respectively and the layers are separated by a distance $d=10$nm.
\label{fig:RxyG}}
\end{figure}

By considering a case $x\rightarrow 0$ and $\alpha_- \rightarrow \infty$, however $x\alpha_-\rightarrow 0$ in which the Hall resistivity is noticeably large, we have
\begin{eqnarray}
\frac{R_H}{R_{H0}}&=&\frac{D\sigma_0}{\beta_0}\frac{\sigma_+\beta_++\sigma_-\beta_-}{(\sigma_++\sigma_-)^2+x^2(\sigma_+\alpha_++\sigma_-\alpha_-)^2}
\nonumber\\
&=&\frac{n_1I_0^2}{4n_\pm I_\pm^2}
\end{eqnarray}
which has a large value at $B$ close to the $B_{S1}$ in which $n^-$ becomes very small. It should be noted that such a huge resistivity does not appear in a 2DEG since the Hall coefficient of a 2DEG at $x=0$ is given by~\cite{Herbut}
\begin{equation}
R_H/R_{H0}=\frac{2\sum_\pm n_{\pm}(I_0/I_\pm)^2}{(\sum_\pm n_\pm I_0/I_\pm)^2 }
\end{equation}
which is a constant at $x\rightarrow 0$. On the other hand, If we consider the case that $\alpha_- \rightarrow \infty$ and $x\alpha_-\rightarrow \infty$ as well, the $R_H$ will be independent of the $x$ value too. In this limit, Eq.~(\ref{eq:1}) takes the following form
\begin{align}
\frac{R_H}{R_{H0}}
=\frac{D\sigma_0}{\beta_0}\frac{\sigma_+\beta_+(x^2\alpha_-^2)+\sigma_-\beta_-(1+x^2\alpha_+^2)}{(\sigma_+(x^2\alpha_-^2))^2+x^2(\sigma_+\alpha_+(x^2\alpha_-^2))^2}
\end{align}

By using the approximation $\beta_- \sim \alpha_-$ in this limit, the term with $\alpha_-^2$ dominates and also $D\rightarrow (1+x^2\alpha_+^2)x^2\alpha_-^2$ and thus we get

\begin{eqnarray}
\frac{R_H}{R_{H0}}
=\frac{\sigma_0 \beta_+}{\sigma_+ \beta_0}=\frac{n_1}{n^+}
\end{eqnarray}

Interestingly, this ratio becomes unity when $B=B_{S1}$. Figure \ref{fig:RxyG} shows the behavior of the normalized $R_{H}/R_{H0}$ as a function of the magnetic field for different values of $x$ for different systems. As shown, the Hall coefficients have a peak in region $B<B_{S1}$ which is the same as one gets in a 2DEG system however it changes for $B>B_{S1}$ and the results depend on the value of $x$. The peak at $B>B_{S1}$ becomes larger by reducing $n_2$ and increases the interlayer correlation. By increasing $x$ value, the shape of the Hall coefficient changes significantly and decreases rapidly as $x$ increases. Contrary to the $R_{xx}$ where the interaction with graphene (2DEG) makes lower (higher) values of the magnetoresistance of layer I, the interlayer interaction always makes $R_H/R_{H0}$ increase compared to its single layer values.

\section{Conclusion}\label{sec4}

In summary, we calculated the longitudinal resistivity and Hall coefficient of the gapless graphene (layer I) in a hybrid structure and compared the results with those of a single layer of graphene in the presence of a tilted magnetic field. We have assumed that the magnetic field is slightly off the electronic planes so that there is a weak perpendicular component.

For a parallel magnetic field, $B_{\bot}=0$ we have analyzed the dependence of the dielectric material which fills the space between the two layers as well as the distance between the layers on the magnetoresistances of layer I and shown that the interlayer interaction plays an important role even at longer distances by using large values of the dielectric constant. Our numerical results show that the resistivity of layer I is significantly modified when $n_2<n_1$. Moreover, the resistivity shows a negative differential magnetoresistance for a small $d$ and for a region in which $B_{||}$ is much smaller than the saturation magnetic field. We have also obtained the reduction of the resistivity when the dielectric constant between two layers increases and obtained the negative differential magnetoresistance in the area where $B_{||}<B_{S1}$ due to the correlation effects induced on layer I from another layer.
We have shown that the $R_{xx}$ in a G-2DEG structure decreases at a small $d$, behaves nonmonotonically as a function of $d$ and the results show in different trend as compared to those in DLG systems.

In the presence of a tilted magnetic field, $B$, the Hall coefficient of layer I shows two peaks at finite $x$ values in which one peak occurs just below the $B_{S1}$ value and another takes place above that. These peaks become sharper at very small $x$ values. Accordingly, a giant like Hall coefficient behavior for layer I can be reached in a hybrid graphene structure. We have shown that the magnetoresistance and the Hall coefficient of a doped graphene layer can be tuned by the electron density and dielectric constants of the materials in a hybrid structure. Our results should be verified by experiments.

\section{Acknowledgment}

R. A. would like to thank the Victoria University of Wellington for its hospitality during the period when the last part of this work was carried out.

\end{document}